# Measuring poverty in India with machine learning and remote sensing

Adel Daoud[1,2,*], Felipe Jordan[3], Makkunda Sharma[4], Fredrik Johansson[2], Devdatt Dubhashi[2], Sourabh Paul[5], and Subhashis Banerjee[6]

1. Institute for Analytical Sociology, Department of Management and Engineering, Linköping University, Norrköping, Sweden.

2. The Division of Data Science and Artificial Intelligence of the Department of Computer Science and Engineering, Chalmers University of Technology, Gothenburg, Sweden

3. UC Santa Barbara, Santa Barbara, CA, USA.

4. Wadhwani AI, Computer Science, Indian Institute of Technology Delhi, India.

5. Humanities & Social Sciences, Indian Institute of Technology Delhi, India.

6. Computer Science and Engineering Indian Institute of Technology Delhi, India.

* Corresponding author: adel.daoud@liu.se

Abstract: In this paper, we use deep learning to estimate living conditions in India. We use both census and surveys to train the models. Our procedure achieves comparable results to those found in the literature, but for a wide range of outcomes.

## 1 Introduction

While country-level data on health and material-living standards—*human development* for short—exist in plentitude for the developing world, within-country data on human development at high spatial and temporal resolution are limited (Atkinson, 2016; Daoud et al., 2016; Deaton, 2015; Halleröd et al., 2013; Jean et al., 2016). Such local data, measuring human development of villages and neighborhoods, are critical for monitoring progress towards the Sustainable Development Goals and enabling tailored public policies to speed up development (Burke et al., 2021). Traditionally, scholars have used either census or household surveys to assess human development (Atkinson, 2016). A *census* is a comprehensive measurement of the material living standard of all individuals in population, yet it is conducted infrequently (usually every 10 years) and collects a small number of characteristics of the target population (Randall and Coast, 2015). In contrast, while household surveys cover a wide range of variables that usually include health outcomes, these surveys have limited reliability for local statistical inference. Increasing the frequency of censuses or surveys would cost governments hundreds of millions of US dollars (Atkinson, 2016). Despite an increase in surveying frequency, scholars would still not have a method for traversing back in time to measure the history of human development.

As reliable satellite imagery have existed since the 1980s (Young et al., 2017), remote surveying using such imagery provides a low-cost, yet reliable alternative to track human development at a fine spatial and temporal resolution (Jean et al., 2016). Twenty years ago, researchers started to explore this avenue by using nightlight luminosity as proxy for human development (Chen and Nordhaus, 2011; Doll et al., 2006; Henderson et al., 2018, 2012; Sutton et al., 2007). Recent advances in the overlap of earth observations (EO) and computer science have made significant progress since then, offering an alternative measurement method that combines daylight- and nightlight satellite images with machine learning (ML) methods to estimate, or *measure*, local characteristics of cities, villages, and neighborhoods (Burke et al., 2021; Daoud and Dubhashi, 2021; Head et al., 2017; Jean et al., 2016; Kino et al., 2021; Yeh et al., 2020). While an EO-ML method still requires ground truth data from census or surveys for training, it tends to outperforms the use of nightlight luminosity in the task of measuring human development (Jean et al., 2016). Nonetheless, existing EO-ML methods are limited in at least two ways.

First, most of the success of existing EO-ML methods have been based on measuring human development in Africa (Burke et al., 2021; Head et al., 2017; Jean et al., 2016; Yeh et al., 2020). A critical question is whether the success of EO-ML methods is tied to cultural, demographic, or economic idiosyncrasies of African data, and thus, how well EO-ML methods generalize to other parts of the world remain to be evaluated.

Second, most of the existing EO-ML methods have been tested on a limited number of cross-sectional outcomes. For example, both Yeh et al.(2020) and Jean et al.(2016) test their method on household income and an asset index that captures household's general material-living standards. Although Head et al.(2017) makes progress in extending the EO-ML method to other dimensions of human development, their study focuses on a limited number of countries (Nepal, Haiti, Nigeria, and Rwanda) and dimensions (e.g., electricity, mobile-phone ownership, child nutrition). Thus, much remains to be done to evaluate how well an EO-ML method can measure other aspects of human development from space.

Relatedly, existing EO-ML methods assume that the definition of the outcome of interest remains constant over time. Yet censuses and surveys do modify their variable definitions to capture temporal variations in the target population (Randall and Coast, 2015). For example, while an individual possessing a mobile phone 20 years ago was an indicator of material wealth, today this product is omnipresent and thus less suitable as an indicator of material wealth (Daoud, 2018; Gordon and Nandy, 2012; Nandy et al., 2016). Correcting for changes in variable definitions remains a challenge for effectively fusing together household surveys in the benefit of training EO-ML models for measuring human development.

If scholars had an EO-ML method that captured human development outside of Africa, they would start making progress towards evaluating the validity and reliability of combining ML and satellite images for measuring human development globally, with high frequency and crisp granularity (Burke et al., 2021; Deaton, 2015). India hosts about 1.4 billion of the world's population, yet it lacks consistent temporal and spatial human-development (Alkire and Seth, 2015; Daoud and Nandy, 2019). With increasing validity and reliability for India, EO-ML methods will become more trustable in measuring the trajectories of human development spatially and temporally. For example, these trajectories will reveal how fast villages or neighbourhoods are lifting out of poverty and ill health, and how well public policies are working in Indian localities.

The aim of our article is to develop EO-ML methods that are capable of measuring human development in selected Indian states across time and space at the village-level. Our aim contributes to addressing the two aforementioned knowledge-gaps. First, we move beyond Africa, and into one of the most populous countries in the world: India. Second, while existing studies develop EO-ML methods for only a few outcomes—most notably income and a household asset index—we tailor and evaluate our EO-ML methods for over 90 outcomes that capture many dimensions of human development. Instead of relying on nightlight luminosity (Xie et al., 2015), our method uses transfer learning based on an asset index we developed from the Indian Census. We find that using our asset index instead of nightlight, achieves superior performance.

To develop our EO-ML method, we require a reliable household-level data source for training our algorithm; a source that can leverage the rich information existing in daytime satellite images (Burke et al., 2021; Rolf et al., 2021). As census data with geographical information system (GIS) is population-wide, we use these data at the village-level to inform our EO-ML method. Additionally, we use the Indian National Family Health Survey (NFHS) for complementary analyses. We combine ML, more specifically deep learning, and Landsat 7 satellite images to measure human development at the village level between 2001 and 2019. For data availability reasons, we limit our analysis to six Indian states (Punjab, Haryana, Uttar Pradesh, Bihar, Jharkhand and West Bengal) that jointly constitutes about 30 percent of the landmass of India.

# 2 Method

## 2.1 Data

Figure 1 provides a map of all of India and our area of study. To reiterate, due to data availability of village-level-administrative data, we restrict our sample to rural areas of six states (Uttar Pradesh, West Bengal, Bihar, Jharkhand, Punjab and Haryana), covering around 218,000 villages, which is about one-third of all Indian villages. Figure 1 includes each included State with its population share of all-India.

[Figure 1 about here]

Our analysis relies on five sources of data: (1) household data from the Indian censuses 2001 and 2011 to primarily measure demographic and material-living conditions, (2) the NFHS from the years 2016 (called NFHS-4) and 2019 (NFHS-5) to measure health outcomes; (3) village-level-administrative boundaries (shapefiles); (4) satellite imagery from the Landsat 7 mission from the years 2001, 2011, 2016 and 2019; and (5) nightlight data provided by the Defense Meteorological Satellite Program's Operational Linescan System (DMSP-OLS).

#### 2.1.1 Census and the National Family Health Survey

From the 2011 census, we create 16 asset indices measuring material-living standards, defined in Table 1. For each index, we calculate the average occurrence of material assets in a village. For example, "electronics" is the average occurrence of "radio, transistor, tv and laptops."[1] We denote the 16-dimensional material-asset vector of outcomes from the census as $Y^c$.

[Table 1 about here]

As shown in Table 2, while the 2001 census has similar definitions for most variables compared to 2011, some of them are measured at a different level of aggregation. This mismatch in aggregation makes it challenging to compare across the two census rounds. For example, in 2011, the village-level data report the percentage of households having a telephone. In comparison, the 2001 data uses a binary variable indicating whether a telephone is available in the village. There is a lack of census documentation on how to translate a continuous proportion value in 2011 to binary value in 2001. Because of this lack, we cannot directly compare these sets of outcomes in the 2001 and 2011 censuses at the village level.

A redeeming aspect of the 2001 Census is that the values for 10 of the 16 variables we construct using 2011-census data are available as a fraction of the population at the tehsil level, one administrative level up from villages. Table 2 shows how these ten variables are constructed from 2001 and 2011 tehsil-level census data. Our indirect method relies on aggregating across Indian administrative levels. India is divided into the following administrative units, starting from the finest to the least coarse: a village comprises around 1000 households; a tehsil covers about 100 villages; a district (with an average population of 1.1 million) has around ten tehsils; and a state captures a set of districts. Our data has 189 districts in six states in Northern and Eastern India. Although a temporal evaluation for some variables is not possible at the finest level of aggregation (village level), it is still possible if we accept an aggregation at the tehsil-level. The method section delineates our aggregation procedure.

[1] We gave each underlying asset equal weight in the aggregation, yet a task for future research is to weight these assets depending on their contribution to human development.

[Table 2 about here]

While the census measures material-living standards, the NFHS captures mainly health outcomes. The NFHS is a multi-round survey providing information on the health of women and children. We collected 93 outcomes that we denote $Y^s$, that is a vector, where the index $s$ stands for "survey." These outcomes are defined in Table 3. As NFHS surveys do not release the latitude and longitude of households, we predict human development at the village level and then aggregate these village-level predictions to the district level for model evaluation.

[Table 3 about here]

### 2.1.2 Landsat 7 satellite images

To map villages on their satellite-image locations, we require information on each village's shapefiles. Once we have mapped a village to its satellite images, we can start extracting satellite-image characteristics of human development for this village, based on its daytime-visual appearance from space. For a village $i$, we denote these characteristics $X_i^d$, as a 3-dimensional tensor (with width, hight, and number of image bands). Then, the collection of $X_i^d$ across time *t* as a 4-dimensional tensor $X_{it}^d$. The index *d* denotes daylight images.

We build a single image for each village *i* and year *t*, using Google Earth Engine's Tier 1 and Tier 2 repositories of Landsat 7 daytime imagery (Gorelick et al., 2017). Tier 1 contains imagery of the highest quality. Tier 2 imagery have the same radiometric standard as that of the Tier 1 imagery, but do not meet Tier 1's geometric specifications. We process these images on Google Earth Engine's servers into batches. In a given batch and year, we use the Quality Assessment Band to remove imagery with more than 5% of cloud coverage, defined as the union of all the 3.36 x 3.36 km squares based on the centroid of the villages. Then, we select four bands for processing: the red, green, and blue soil reflectance bands, plus the panchromatic top-of-the-atmosphere band.

The selected images are processed with the C-correction Teillet method to smooth the effect of topography in the imagery (Ekstrand, 1996; Riano et al., 2003; Richter et al., 2009), which has been shown to improve the capacity of EO-ML methods to distinguish among land cover classes (Khatami et al., 2016). The selected imagery is then ordered from the one with the highest level of Normalized Difference Vegetation Index (NDVI) within the area of interest to the one with the lowest NDVI. We then build a mosaic across the area of interest through a recursive method that selects, from the first imagery, all the valid pixels, and then moves to the second to fill pixels that were covered by clouds or saturated in the first image. This method continues down the list until all pixels are filled or the end of the list is reached.

The use of year-round composites allows us to maximize the probability of obtaining data for all pixels in subtropical regions with cloudy wet seasons, and the use of NDVI as a quality measurement maximizes the chances of observing standing crops in agricultural lands, which is likely to help the model differentiate agricultural land from bare soil. This composite also enables us to fill the gaps produced by the failure of the Scan Line Corrector found in images collected after May 31, 2003. Finally, we use a simple MRA pansharpening method to combine the RGB 30-meter resolution bands with the 15-meter resolution panchromatic band and create our final 15-meter resolution imagery (Vivone et al., 2015). Each band is normalized to have a mean of zero and a standard deviation of one across all image samples $X_{it}^d$. When unpacking the batch, the result is a sequence of 224x224x3 input tensors, each of which is associated to a single village-year, covering 11.3 square kilometers. The 224x224 is the height and width of the geographical area measured in pixels. Because of pansharpening, each pixel has a size of 15 meters by 15 meters. Each pixel has three bands covering the colors red, green and blue. We selected the size of the geographical area based on established deep-learning-model architectures (Krizhevsky et al., 2017).

While satellite images capture how a village's material living standards appear from the sky during daylight $X^d$, we use nightlight data $X^\eta$ to quantify how much luminosity a village emits during the night. Each pixel contains one band (luminosity value) about $X^\eta$. The more luminosity emitted, the higher the material-living standards tend to be (Chen and Nordhaus, 2011; Doll et al., 2006). We collect nightlight data from DMSP-OLS. The night light data is available in 30 arc second grids (about 800 meters at the latitude of New Delhi), spanning - 180° to 180° longitude and -65° to 75° latitude. Each 30 arc second grid cell is mapped to integer values from 0 to 63, where 63 corresponds to the highest night light intensity.

### 2.2 Method

We will use all the aforementioned data to train a set of EO-ML deep-models $f$ at the village level and always using daylight $X^d$ as input and predict the census-measured $Y^c$ human development outputs. Via transfer learning, we will then predict the survey-measured $Y^s$ outputs. For the baseline model $f_b$, we will also use $X^d$ as input but nightlight data $X^\eta$ as output, in a transfer learning step, and then, either predict survey- $Y^s$ or census-measured $Y^c$ human development.

#### 2.2.1 Deep learning

We used a variety of deep-learning architectures, denoted as a set of functions $f$. To evaluate model dependency in our experiments, we used the following set of functions: ResNet-18 ($f_{18}$), ResNet-34 ($f_{34}$), ResNet-50 ($f_{50}$), VGG-16 ($f_{16}$) convolutional neural network (He et al., 2016; Simonyan and Zisserman, 2015) (CNN). All models are pre-trained on ImageNet (Deng et al., 2009), and then tuned for our regression tasks to predict $Y^c$. We replace the last fully connected layer of each pre-trained network, from one of dimensions 512 × 1000, to a randomly initialized layer of dimensions 512 × 16.

We use the Adam optimizer with learning rate $10^{-3}$ and a batch size of 64. To enhance model-training performance, we compute and use the normalized band-wise mean and standard deviation of our dataset instead of the mean and variance of the ImageNet data.

### 2.2.1 Baseline model: nightlight

As previous research has shown that models that combine daylight imagery $X^d$ and nightlight imagery $X^\eta$ for predicting material-living conditions outperform models that do not use this combination (Xie et al., 2015), and thus, we use this combination as a baseline model, $f_b$. As Figure 2 shows, our baseline consists of first training an ML model to predict nightlight values from daytime satellite imagery. We identify a set of relevant nightlight cells by using the latitude and longitude of the centroid of the daytime satellite images. We perform the regression with a train-test split of 8:2.

[Figure 2 about here]

We then use the nightlight model via a transfer-learning procedure to predict a 16-dimensional vector representing the household level material-assets in a village. The output of the nightlight model at the last hidden layer serves as a feature vector which provides a 512-dimensional representation of the village's level of human development, captured by the night light as a proxy. We use this for transfer learning to the material-asset vectors. We use the 224 × 224×3 daytime images from Landsat 7 as input to the baseline $f_b$ to get a 512-dimensional representation and design a neural network with a single fully connected layer with rectified linear activation to do a regression of the asset vectors $Y^c$ on the 512-dimensional representation.

Moving beyond the baseline model, we train ML models to learn the 16-dimensional asset vector directly from daytime imagery without using the nightlight as an intermediate training step.

### 2.2.1 Transfer Learning from census to survey

Our research design relies on two assumptions, First, a key assumption of our research design is that satellite images capture human development from the sky. That is, $X^d$ contains predictive information about the asset outputs $Y^c$. In other words, we are assuming that the asset indicators are correlated with the observable features in the daytime satellite images like proportion of built-up area, road area and road types, density and type of housing, water bodies, forest cover and green areas, etc.

Our second assumptions is that if such correlations exist between $X^d$ and $Y^c$, we can use transfer learning to learn outcomes $Y^s$ that may not directly correlate $X^d$. We investigate whether the direct regression model trained to predict the asset indicators can be used for transfer learning of other socio-economic indicators $Y^s$for literacy, health and demographics (Daoud et al., 2019; Kino et al., 2021). With transfer learning, we can expand the dimensions of human development that can be studied, even in small datasets that do not offer a large-enough training set (Zhuang et al., 2021). This augmentation occurs by combining two tasks. First train on the task $f: X^d \rightarrow Y^c$, and then modify the last layers of $f$ the deep-learning model before initiating the second task. Call this modified model $f^*$ and now the second task is to find a mapping between input and output such as, $f^*: X^d \rightarrow Y^s$. Via transfer learning, $f^*$ benefits from the training experience of $f$.

While NFHS-4 and NFHS-5 $Y^S$ data are available at the district level, census $Y^C$ is available at the village level. There are a total of 189 districts in the six north Indian states. For transfer learning, we aggregate the asset model regression output. That is, we pass the village Landsat image to our model and extract the layer just before the final prediction layer of the 16-dimensional outputs. Then, we average this layer across all villages in each district weighting by villages' populations. Using this averaged layer as inputs, we train a neural network with two fully connected layers with rectified linear activation to do a regression on the NFHS-4 outcomes.

We assess the performance of the proposed method, that uses a 16-dimensional census outcome vector to train the model that extract the most relevant features from the imagery, with the current standard in the literature, which uses nighttime light Data instead (Jean et al., 2016).

### 2.2.2 Temporal evaluation

To evaluate the performance of our EO-ML models across time, we perform experiments where we take village daylight-satellite images from 2011, $X^d_{2011,i}$ pass them through our models, $f\left(X^d_{2011,i}\right) = \hat{Y}^c_{2011,i}$ for training. Then, for evaluation we use held-out 2001-census data, $Y^c_{2001,i}$ and $X^d_{2001,i}$. That is, $f\left(X^d_{2001,i}\right) = \hat{Y}^c_{2001,i}$. As for 10 of 16 outcomes the census-variable definitions have changed between 2001 to 2011 at village level, we conduct the temporal evaluation at the tehsil level, where the definitions match. We aggregate the analysis to the tehsil level, indexed by $h$, by weighting $w_i$each village prediction $\hat{Y}^c_{2011,i}$ using each village's population size. This aggregation creates a tehsil prediction vector, $\sum_i w_i \hat{Y}^c_{2011,i} = \hat{Y}^c_{2011,h}$, and a tehsil level outcome for evaluation $\sum_i w_i Y^c_{2001,i} = Y^c_{2001,h}$. Thus, the target loss we aim to minimize is the sum of squares over all tehsils, that is, $\sum_h \left(\hat{Y}^c_{2011,h} - Y^c_{2001,h}\right)^2$.

A key problem is a lack of stationarity of the outcome and input (satellite) distributions between the years 2001 and 2011. Stationarity is a process that maintains its mean, variance, rank order, and autocorrelation structure over time. In our case, while the input distribution is reasonably stationary—that is human development is a slow process that satellite images capture—the outcome distribution for some variables changes rapidly. This shift in distributions for some outcomes challenges an approach that aims to directly evaluate ML models across time, and thus, we resort to an indirect method to handle this challenge.

We tested three distribution transformation, call them $g()$ to align 2001 to 2011 ground truth distributions. Because of socio-economic changes between these two years, the outcome distributions have shifted in the gap of 10 years, making it more difficult for any model to predict temporal change. The first four central moments (mean, variance, skewness, and kurtosis) of these two distributions do not match. To mitigate this problem, we applied the following transformations on the ground truth before model evaluation:

- *Simple transform*, matches the mean and variance of the 2001 ground truth census data to mean and variance of 2011 census data before evaluation.

- *Histogram matching*, transforms 2001 census data by matching histograms of the 2001 census and 2011 census at 10 bins for each variable.

- *Linear optimal transport*, learns a linear optimal transport from 2001 census to 2011 census on the training data, and applied it to 2001 census test ground truth before evaluating. Add reference

Thus, the target loss that we minimize after each of these transformations is the sum of squares over all tehsils, $\sum_h \left(\hat{Y}^c_{2011,h} - g(Y^c_{2001,h})\right)^2$.

### 2.2.3 Aggregations

In summary, for all levels of aggregation and experiments, the satellite-input data is always collected at the village level $X^d_{it}$. Yet our analysis relies on different aggregation levels, because of either mismatch in definition at the village level or missing latitude-longitude information for outcomes $Y_i$. First, all census-based cross-sectional results rely on village-level data *i*, and thus, do not use any aggregation. Second, for all NFHS experiments we conduct the evaluation at the district level. While the ground-truth data $Y^s_i$ is collected at the village-level, we aggregate to the district level *d*. Third, for the temporal analysis, we conduct the model evaluation at the tehsil level.

# 3 Results

## 3.1 Cross-sectional and transfer learning results

We trained four ML models ResNet-18, ResNet-34, ResNet-50, and VGG-16. As shown by panel (a) in Figure 3, although all models had comparable model performance, ResNet-18 performed slightly better for 13 of the 16 outcomes. ResNet-34 trails the ResNet-18 performance, and ResNet-34 performs better on water-related outcomes, with $R^2$ in the range of 0.3 and 0.4. Henceforth, we only present results from ResNet-34, as this is the best overall-performing model.[2] ResNet-34 is performing closely to the performance of ResNet-18 on 13 of 16 outcomes, but better than ResNet-18 on the last three outcomes.

[2] Other model results are available upon request.

[Figure 3 about here]

Our ResNet-34 (henceforth referred to as $f_{34}$) model's average $R^2$-performance, across all 16 outcomes using census 2011, is 0.5, with a standard deviation of 0.12. As panel (a) in Figure 3 shows, our $f_{34}$ performance ranges from as high as $R^2 = 0.69$ (permanent house) to a low of $R^2 = 0.27$ (water-treated). All our models perform well on outcomes that has a physical appearance from the sky—such as housing quality—and it tends to struggle with outcomes that merely correlate with outcomes appearing from the sky (e.g., a household's water-quality access correlates with housing quality).

For the results in panel (a) Figure 3, our models are directly learning to associate the daylight-image input, $X_i^d$ to the census-material-living-standards outputs $Y_i^c$. That is, no aggregation or transfer learning is used.

In panel (b) Figure 3 (census results) and all results in Figure 4 (NFHS results), our $f_{34}$ rely on transfer learning. The left graph shows the results of the $f_{34}$ when using the 16-dimensional census variables as the output of the first step of the transfer learning. As a benchmark, we present the results of the $f_{34}$ based on nightlight images as the output of the first step of the transfer learning in the right graph (Jean et al., 2016). Model evaluation is conducted at the district level. As described in the method section, we conduct transfer learning in two steps. First, we use $f_{34}$ that learned to predict $Y^c$ (i.e., the results from panel a Figure 1) and modify the last layer in $f_{34}$ so it can predict new outcomes from the census $Y^{c*}$ and NFHS $Y^s$, respectively. We call this modified model $f_{34}^*$. Second, we train $f_{34}^*$ using $Y^{c*}$ and $Y^s$ respectively. While panel (b) in Figure 3 shows the results for $Y^{c*}$, panel (a) in Figure 4 shows the result for $Y^s$.

The key innovation here is that $Y^{c*}$ and $Y^s$ contains outcomes that are distal from what the combination of satellite-images and $f_{34}$ can be expected to predict. For example, as previously mentioned, predicting housing quality from satellite images is not an unreasonable task, because roofs, roads, and yards are directly observable from the sky; but predicting literacy rate or religious affiliation is a distal outcome that is not directly observable from satellite images. Despite the outcomes contained in $Y^{c*}$(such as the rate of literacy, working population, and the religious caste) are not directly observable from the sky, our $f_{34}^{*}$ produces acceptable performance for several of $Y^{c*}$.

Thus, panel (b) in Figure 3 shows that our model $f_{34}^{*}$ can predict not only material-living standards but also demographic characteristics with reasonable performance. The model performs best on measuring the share of *scheduled tribe* ($R^2$=0.49), followed by *literacy rate* ($R^2$=0.34), working population ($R^2$=0.15), and scheduled caste ($R^2$=0.1).

As shown in panel (c) in Figure 3, Comparing a model that use nightlight luminosity in the transfer-learning step instead of our $f_{34}^{*}$ that uses the 16-dimensional vector for transfer learning, we observe that our $f_{34}^{*}$ performs the best. A nightlight-transfer model has the following performance: *literacy rate* ($R^2$=0.21), *scheduled tribe* ($R^2$=0.18), working population ($R^2$=0.08), and scheduled cast ($R^2$=0.06). That our transfer-learning approach performs better than nightlight, indicate that the combination daylight-asset conditions contains more information to predict distal demographic characteristic. As nightlight only captures black-and-white luminosity changes, it is not as information rich as daylight-asset captures in characterizing human development.

Similarly, although $Y^s$ contains mainly health outcomes, $f_{34}^{*}$ has the capacity to predict several of them reasonably well. Figure 4 shows the results for transfer-learning based on NFHS outcomes $Y^s$. Panel (a) shows that our model $f_{34}^{*}$ identifies sufficient signal to predict a variety of health-related outcomes. The histogram shows the frequency outcomes over levels of $R^2$. Of the 93 outcomes, 27 had a score of $R^2 \geq 0.5$. As shows in panel (b), the five top-scoring variables in $Y^s$are the following: *precent of women overweight*, *households with clean fuel*, *share of population below age 15*, *access to condoms for birth control*, and *birth with caesarean section*. The bottom-five outcomes in $Y^s$ are the following: *vaccination against measles*, *men with high blood sugar*, diarrhea treated with zinc, *women with anemia*, and *women with high blood pressure* (BP). These variables have all negative $R^2$, which means that the predictions of $f_{34}^{*}$ are worse than just using the sample mean for each $Y^s$.

[Figure 4 about here]

## 3.2 Temporal results

Our $f_{34}^*$ trained on the 16-vector census 2011 is able to predict temporal changes, targeting the census 2001 and both NFHS surveys in 2015 and NFHS in 2019. Panel (a), Figure 5, shows the census-2001 results. As described in the Method section, besides the non-transformed (original) data, we use three procedures to align the outcome distributions. Using no transformation, the model produces poor results. The worst performing outcome is has-phone, the proportions of phones in villages, with an $R^2 = -236$. The model produces negative $R^2$ for *electronics* ($R^2 = -0.1$), *banking-services availability* ($R^2 = -2.7$), and *no-assets* ($R^2 = -0.5$). As previously mentioned, negative $R^2$ means that the model is performing worse than a prediction using the sample mean.

[Figure 5 about here]

While the simple-transform algorithm produces uneven results with negative and positive $R^2$, the best performing transformations are histogram matching and linear-optimal matching. Based on these two transformations, our model $f_{34}^*$ estimates the following outcomes with $R^2 \geq 0.5$: *oil-like* (having energy source from kerosene/other oil), *electric-like* (having energy source from grid/solar), *bathroom-within* (having bathroom within premises), and *cook-fuel-processed* (having LPG/electric stove). Outcomes that $f_{34}^*$ tends to estimate less precisely are *electronics* (having possession of radio/transistor/tv/laptop) and *has-phone* (having possession of land-line/mobile/both). One reason for this less precise predictions is that electronic items have become omnipresent, even among the poorer segments of the population.

One reason to why histogram matching and linear-optimal transform perform better than simple transform, is that simple transform only aligns the two first central moments (mean and variance), while the other two aligns the 2001 and 2011 census-distributions across different distribution characteristics. The poor performance of $f_{34}^*$ using non-transformed data is because the model is unable to predict the mean (levels of living conditions), even if the 2001 and 2011 distributions retain their rank order across village,. This poor prediction is because the outcomes have changed faster than the input (satellite) features. For example, while having access to mobile phone in 2001 was considered a luxury item, in 2011 and onwards, mobile phones are owned by even the poorest households.

Panel (b), Figure 5, shows how our $f_{34}^{*}$, trained on the census-2011 data, but now transferred to predicting 93 health outcomes in NFHS-4 (year 2015), using three transformations. The first row shows the cross-sectional result presented earlier, as a point of references. Our model performs comparably well across the three transformations. Panel (c) shows the best and worst performing outcomes. Similar to the cross-sectional results, the following comes out on top with $R^2$ around 0.6: *precent of women overweight*, *share of population below age 15*, and *birth with caesarean section*. Outcomes such as *woman-underaged marriage* and *women with secondary education* are also performing in that $R^2$ range. Health outcomes with negative $R^2$ are, for example, *women with high blood pressure*, *vaccinated against measles,* and *anemic women*. Of all 93 health outcomes, histogram matching produces 69 and linear-optimal transport produces 70 with positive $R^2$.

Panel (c), Figure 5, shows the performance of $f_{34}^{*}$ to predicting NFHS-5 (year 2019) health outcomes. Here, we rely on double-transfer learning. The first transfer step is that $f_{34}^{*}$ predicts NFHS-4 outcomes. Using the first step, now $f_{34}^{*}$ is modified into $f_{34}^{**}$—that is, the last layer in the ResNet-34 architecture is changed for predicting NFHS-5 outcomes. In NFHS-5, we focus on outcomes related to child health. On top of the double-transfer, we also apply the four procedures of distribution transforms. Regardless of transformations, $f_{34}^{**}$ performs well in predicting mother's access to antenatal care, mothers who consumed iron folic acid for 100 days or more when they were pregnant (%), and share of children underweight.

## 4 Discussion

The global community has committed to ambitious targets articulated in the Sustainable Development Goals (Burke et al., 2021; Daoud et al., 2016; Halleröd et al., 2013). Although many governments globally are vigilant in implementing public policies to improve human development for their populations (Conklin et al., 2018; Coutts et al., 2019; Ponce et al., 2017; Shiba et al., 2021), policymakers lack reliable methods to monitor the effects of their policies at a sufficiently granular level over time and space (Daoud, 2015). To tackle this lack, scholars are creating innovative methods that capitalizes on the predictive accuracy of ML and the visual granularity supplied by EO (Burke et al., 2021; Daoud and Dubhashi, 2020; Rolf et al., 2021). As most of these EO-ML methods focus on Africa, our article creates a comparable method for India. That is this article's first contribution: while existing EO-ML methods for Africa cover human development for roughly one-seventh of the world population, our method makes progress towards covering an additional one-seventh.

A second contribution of is that our EO-ML method uses outcome-distribution transformations to better estimate temporal change. Because some dimensions of human development change faster—an outcomes such as *access to mobile phones*—than the material shape of neighborhoods as exhibited in satellite images, EO-ML method can struggle in estimating temporal change (Young et al., 2017). Our analysis shows that accessible transformations such as histogram matching or linear-optimal transform boosts accuracy by several factors.

A third contribution is that our results show that using a census provides a better leverage than nightlight luminosity for transfer learning. While nightlight luminosity is a frequently used complementary data source to estimate human development (Henderson et al., 2012; Xie et al., 2015), our experiments show that a 16-dimensional asset vector performs better. Using this asset vector as a leverage, our EO-ML method transferred, with noteworthy accuracy, to estimate a myriad of health outcomes as measured by NFHS.

Of the 93 health outcomes and for all transformations our experiments evaluated, about 70 had positive $R^2$ values, and the best transformation (i.e., linear optimal transport) produced values $R^2 > 0.5$ for 23 outcomes. Conversely, as the temporal results show, nightlight-based transfer produced consistently less competitive $R^2$ values. Thus, our EO-ML method enables scholars and policymakers to measure health outcomes that are not directly observable from the sky. For example, our top-performing outcomes—with $R^2$ around 0.6—are *precent of women overweight*, *share of population below age 15*, and *birth with caesarean section*, *woman-underaged marriage* and *women with secondary education*.

Improving maternal and child health outcomes are integral part of the Sustainable Development Goals. Although the Indian economy is growing and Indian governments are improving their public-health and anti-poverty policies, much remains for pulling millions out of poverty in the last several decades (Alkire and Seth, 2015; Drèze and Sen, 2013; Reddy and Daoud, 2020; Thorat et al., 2017). For example, India has halved its population-poverty rate from 45.3% (head-count ratio) in 1993 to 21.9% in 2012, yet about 54 million people still live in extreme poverty and with ill-health. To calibrate public policies, policymakers require geo-temporal data for efficient policy targeting (McBride and Nichols, 2016). Our EO-ML method is one critical piece for enabling such efficient targeting.

## 5 References

Alkire, S., Seth, S., 2015. Multidimensional Poverty Reduction in India between 1999 and 2006: Where and How? World Development 72, 93–108. https://doi.org/10.1016/j.worlddev.2015.02.009

Atkinson, T., 2016. Monitoring Global Poverty: Report of the Commission on Global Poverty. The World Bank. https://doi.org/10.1596/978-1-4648-0961-3

Burke, M., Driscoll, A., Lobell, D.B., Ermon, S., 2021. Using satellite imagery to understand and promote sustainable development. Science 371, eabe8628. https://doi.org/10.1126/science.abe8628

Chen, X., Nordhaus, W.D., 2011. Using luminosity data as a proxy for economic statistics. Proceedings of the National Academy of Sciences 108, 8589–8594. https://doi.org/10.1073/pnas.1017031108

Conklin, A.I., Daoud, A., Shimkhada, R., Ponce, N.A., 2018. The impact of rising food prices on obesity in women: a longitudinal analysis of 31 low-income and middle-income countries from 2000 to 2014. International Journal of Obesity 1. https://doi.org/10.1038/s41366-018-0178-y

Coutts, A., Daoud, A., Fakih, A., Marrouch, W., Reinsberg, B., 2019. Guns and butter? Military expenditure and health spending on the eve of the Arab Spring. Defence and Peace Economics 30, 227–237. https://doi.org/10.1080/10242694.2018.1497372

Daoud, A., 2018. Unifying Studies of Scarcity, Abundance, and Sufficiency. Ecological Economics 147, 208–217. https://doi.org/10.1016/j.ecolecon.2018.01.019

Daoud, A., 2015. Quality of Governance, Corruption, and Absolute Child Poverty in India. Journal of South Asian Development 10, 1–20.

Daoud, A., Dubhashi, D., 2021. Melting together prediction and inference. Observational Studies 7, 1–7.

Daoud, A., Dubhashi, D., 2020. Statistical modeling: the three cultures. arXiv:2012.04570 [cs, stat].

Daoud, A., Halleröd, B., Guha-Sapir, D., 2016. What Is the Association between Absolute Child Poverty, Poor Governance, and Natural Disasters? A Global Comparison of Some of the Realities of Climate Change. PLOS ONE 11, e0153296. https://doi.org/10.1371/journal.pone.0153296

Daoud, A., Kim, R., Subramanian, S.V., 2019. Predicting women's height from their socioeconomic status: A machine learning approach. Social Science & Medicine 238, 112486. https://doi.org/10.1016/j.socscimed.2019.112486

Daoud, A., Nandy, S., 2019. Implications of the Politics of Caste and Class for Child Poverty in India. Sociology of Development 5, 428–451. https://doi.org/10.1525/sod.2019.5.4.428

Deaton, A., 2015. The Great Escape: Health, Wealth, and the Origins of Inequality, Reprint edition. ed. Princeton University Press, Princeton, NJ.

Deng, J., Dong, W., Socher, R., Li, L.-J., Kai Li, Li Fei-Fei, 2009. ImageNet: A large-scale hierarchical image database, in: 2009 IEEE Conference on Computer Vision and Pattern Recognition. Presented at the 2009 IEEE Conference on Computer Vision and Pattern Recognition, pp. 248–255. https://doi.org/10.1109/CVPR.2009.5206848

Doll, C.N.H., Muller, J.-P., Morley, J.G., 2006. Mapping regional economic activity from night-time light satellite imagery. Ecological Economics 57, 75–92. https://doi.org/10.1016/j.ecolecon.2005.03.007

Drèze, J., Sen, A., 2013. An Uncertain Glory: India and its Contradictions. Penguin, London.

Ekstrand, S., 1996. Landsat TM-based forest damage assessment: correction for topographic effects. Photogrammetric Engineering and Remote Sensing 62, 151–162.

Gordon, D., Nandy, Shailen., 2012. Measuring child poverty and deprivation, in: Minujin Z., Alberto., Nandy, Shailen. (Eds.), Global Child Poverty and Well-Being: Measurement, Concepts, Policy and Action. Policy Press, Bristol, UK; Chicago, IL, pp. 57–101.

Gorelick, N., Hancher, M., Dixon, M., Ilyushchenko, S., Thau, D., Moore, R., 2017. Google Earth Engine: Planetary-scale geospatial analysis for everyone. Remote Sensing of Environment, Big Remotely Sensed Data: tools, applications and experiences 202, 18–27. https://doi.org/10.1016/j.rse.2017.06.031

Halleröd, B., Rothstein, B., Daoud, A., Nandy, S., 2013. Bad governance and poor children: a comparative analysis of government efficiency and severe child deprivation in 68 low-and middle-income countries. World Development 48, 19–31.

He, K., Zhang, X., Ren, S., Sun, J., 2016. Deep Residual Learning for Image Recognition. Presented at the Proceedings of the IEEE Conference on Computer Vision and Pattern Recognition, pp. 770–778.

Head, A., Manguin, M., Tran, N., Blumenstock, J.E., 2017. Can Human Development be Measured with Satellite Imagery?

Henderson, J.V., Squires, T., Storeygard, A., Weil, D., 2018. The Global Distribution of Economic Activity: Nature, History, and the Role of Trade. Q J Econ 133, 357–406. https://doi.org/10.1093/qje/qjx030

Henderson, J.V., Storeygard, A., Weil, D.N., 2012. Measuring Economic Growth from Outer Space. American Economic Review 102, 994–1028. https://doi.org/10.1257/aer.102.2.994

Jean, N., Burke, M., Xie, M., Davis, W.M., Lobell, D.B., Ermon, S., 2016. Combining satellite imagery and machine learning to predict poverty. Science 353, 790–794. https://doi.org/10.1126/science.aaf7894

Khatami, R., Mountrakis, G., Stehman, S.V., 2016. A meta-analysis of remote sensing research on supervised pixel-based land-cover image classification processes: General guidelines for practitioners and future research. Remote Sensing of Environment 177, 89–100. https://doi.org/10.1016/j.rse.2016.02.028

Kino, S., Hsu, Y.-T., Shiba, K., Chien, Y.-S., Mita, C., Kawachi, I., Daoud, A., 2021. A scoping review on the use of machine learning in research on social determinants of health: Trends and research prospects. SSM - Population Health 15, 100836. https://doi.org/10.1016/j.ssmph.2021.100836

Krizhevsky, A., Sutskever, I., Hinton, G.E., 2017. ImageNet classification with deep convolutional neural networks. Commun. ACM 60, 84–90. https://doi.org/10.1145/3065386

McBride, L., Nichols, A., 2016. Retooling Poverty Targeting Using Out-of-Sample Validation and Machine Learning. World Bank Econ Rev. https://doi.org/10.1093/wber/lhw056

Nandy, S., Daoud, A., Gordon, D., 2016. Examining the changing profile of undernutrition in the context of food price rises and greater inequality. Social Science & Medicine 149, 153–163. https://doi.org/10.1016/j.socscimed.2015.11.036

Ponce, N., Shimkhada, R., Raub, A., Daoud, A., Nandi, A., Richter, L., Heymann, J., 2017. The association of minimum wage change on child nutritional status in LMICs: A quasi-experimental multi-country study. Global Public Health 13, 1–15. https://doi.org/10.1080/17441692.2017.1359327

Randall, S., Coast, E., 2015. Poverty in African Households: the Limits of Survey and Census Representations. The Journal of Development Studies 51, 162–177. https://doi.org/10.1080/00220388.2014.968135

Reddy, S.G., Daoud, A., 2020. Entitlements and Capabilities, in: Martinetti, E.C., Osmani, S., Qizilbash, M. (Eds.), The Cambridge Handbook of the Capability Approach. Cambridge University Press, Cambridge.

Riano, D., Chuvieco, E., Salas, J., Aguado, I., 2003. Assessment of different topographic corrections in Landsat-TM data for mapping vegetation types (2003). IEEE Transactions on Geoscience and Remote Sensing 41, 1056–1061. https://doi.org/10.1109/TGRS.2003.811693

Richter, R., Kellenberger, T., Kaufmann, H., 2009. Comparison of Topographic Correction Methods. Remote Sensing 1, 184–196. https://doi.org/10.3390/rs1030184

Rolf, E., Proctor, J., Carleton, T., Bolliger, I., Shankar, V., Ishihara, M., Recht, B., Hsiang, S., 2021. A generalizable and accessible approach to machine learning with global satellite imagery. Nat Commun 12, 4392. https://doi.org/10.1038/s41467-021-24638-z

Shiba, K., Daoud, A., Hikichi, H., Yazawa, A., Aida, J., Kondo, K., Kawachi, I., 2021. Heterogeneity in cognitive disability after a major disaster: A natural experiment study. Science Advances. https://doi.org/10.1126/sciadv.abj2610

Simonyan, K., Zisserman, A., 2015. Very Deep Convolutional Networks for Large-Scale Image Recognition. arXiv:1409.1556 [cs].

Sutton, P.C., Elvidge, C.D., Ghosh, T., 2007. Estimation of Gross Domestic Product at Sub-National Scales using Nighttime Satellite Imagery. International Journal of Ecological Economics & Statistics 8, 5–21.

Thorat, A., Vanneman, R., Desai, S., Dubey, A., 2017. Escaping and Falling into Poverty in India Today. World Development 93, 413–426. https://doi.org/10.1016/j.worlddev.2017.01.004

Vivone, G., Alparone, L., Chanussot, J., Dalla Mura, M., Garzelli, A., Licciardi, G.A., Restaino, R., Wald, L., 2015. A Critical Comparison Among Pansharpening Algorithms. IEEE Transactions on Geoscience and Remote Sensing 53, 2565–2586. https://doi.org/10.1109/TGRS.2014.2361734

Xie, M., Jean, N., Burke, M., Lobell, D., Ermon, S., 2015. Transfer Learning from Deep Features for Remote Sensing and Poverty Mapping. arXiv:1510.00098 [cs].

Yeh, C., Perez, A., Driscoll, A., Azzari, G., Tang, Z., Lobell, D., Ermon, S., Burke, M., 2020. Using publicly available satellite imagery and deep learning to understand economic well-being in Africa. Nature Communications 11, 2583. https://doi.org/10.1038/s41467-020-16185-w

Young, N.E., Anderson, R.S., Chignell, S.M., Vorster, A.G., Lawrence, R., Evangelista, P.H., 2017. A survival guide to Landsat preprocessing. Ecology 98, 920–932. https://doi.org/10.1002/ecy.1730

Zhuang, F., Qi, Z., Duan, K., Xi, D., Zhu, Y., Zhu, H., Xiong, H., He, Q., 2021. A Comprehensive Survey on Transfer Learning. Proceedings of the IEEE 109, 43–76. https://doi.org/10.1109/JPROC.2020.3004555

# 6 Tables and Figures

## 6.1 Figures

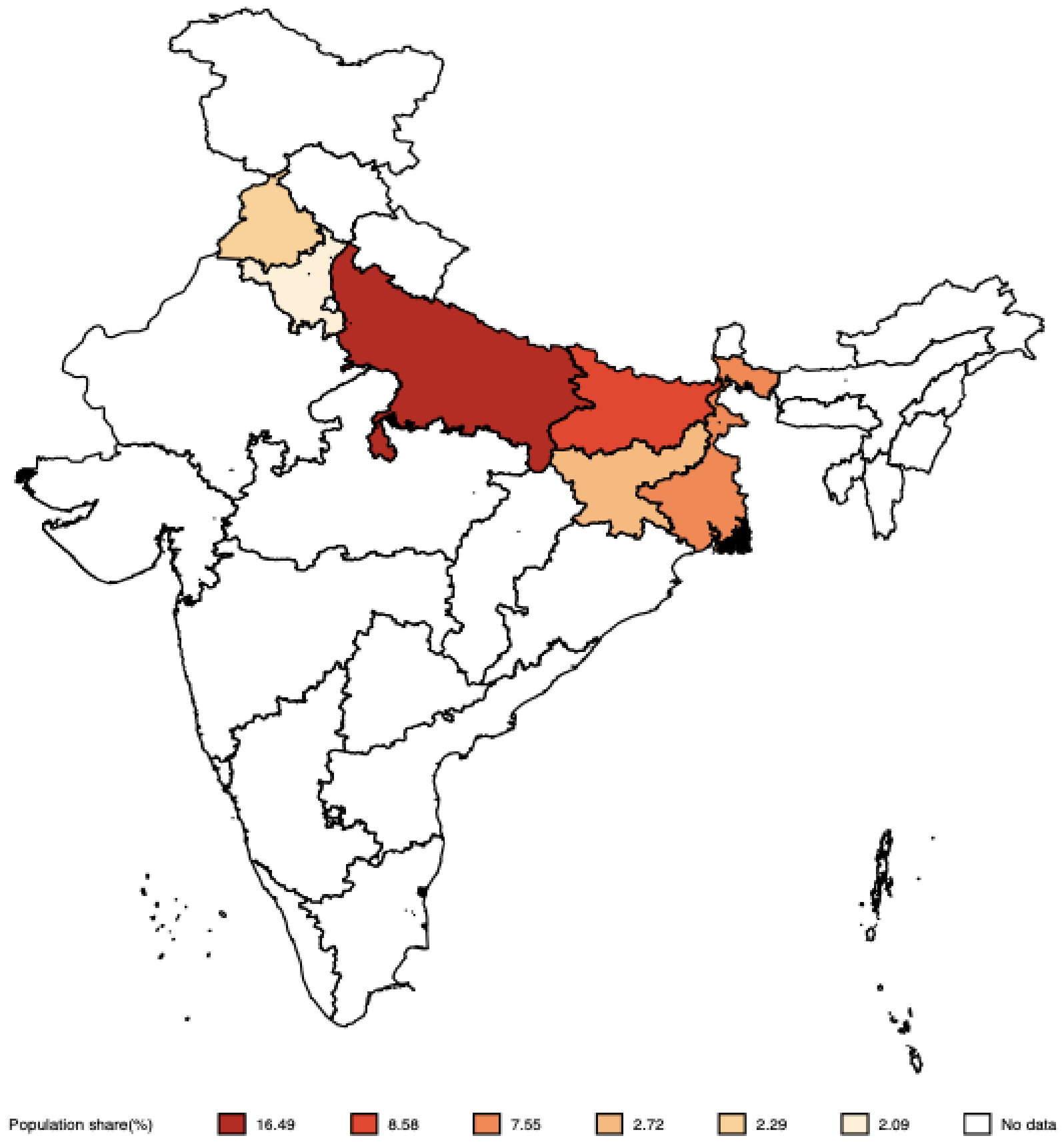

*Figure 1: Scope of the study demographically and geographically*

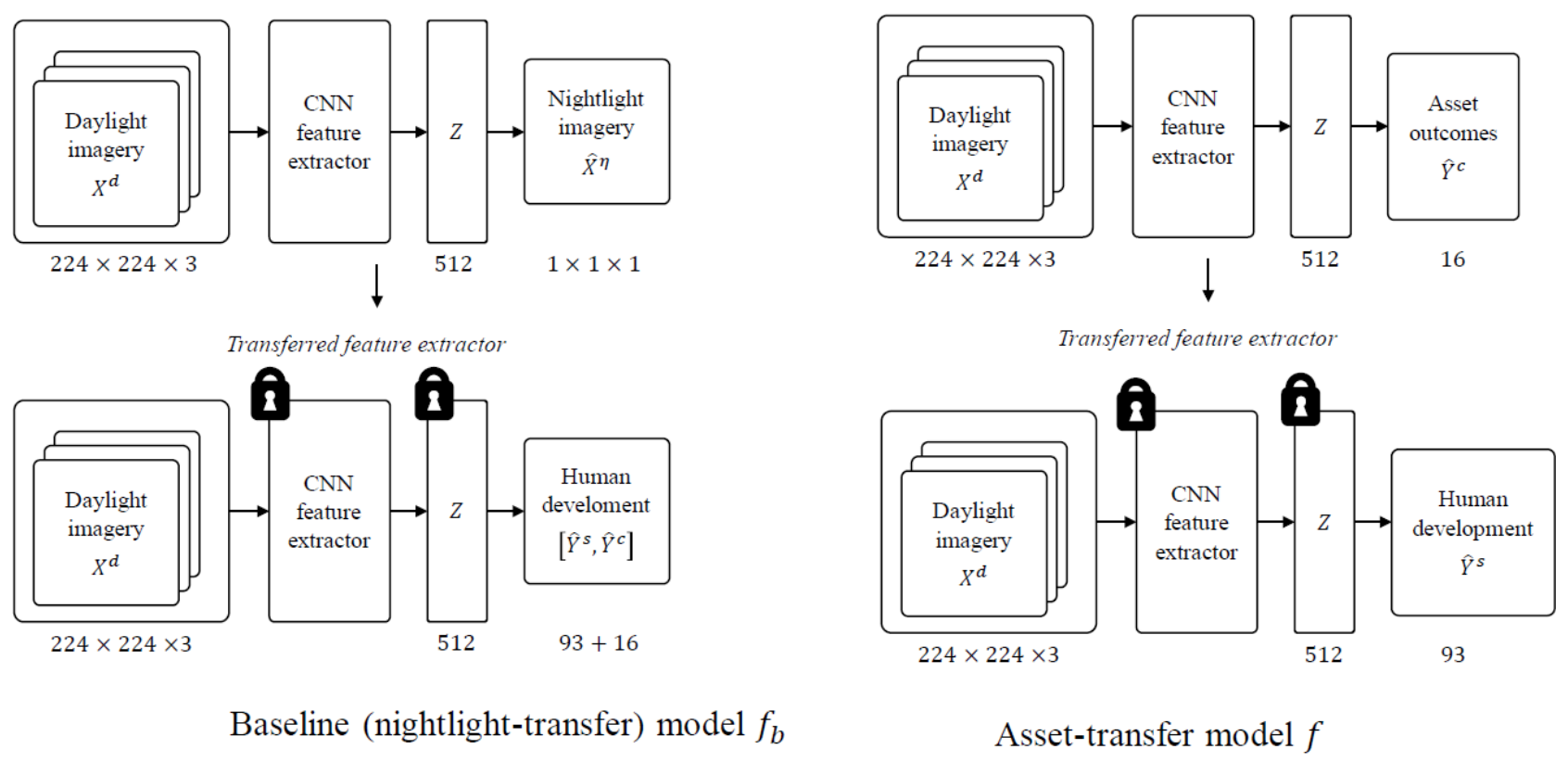

*Figure 2: Deep-learning models*

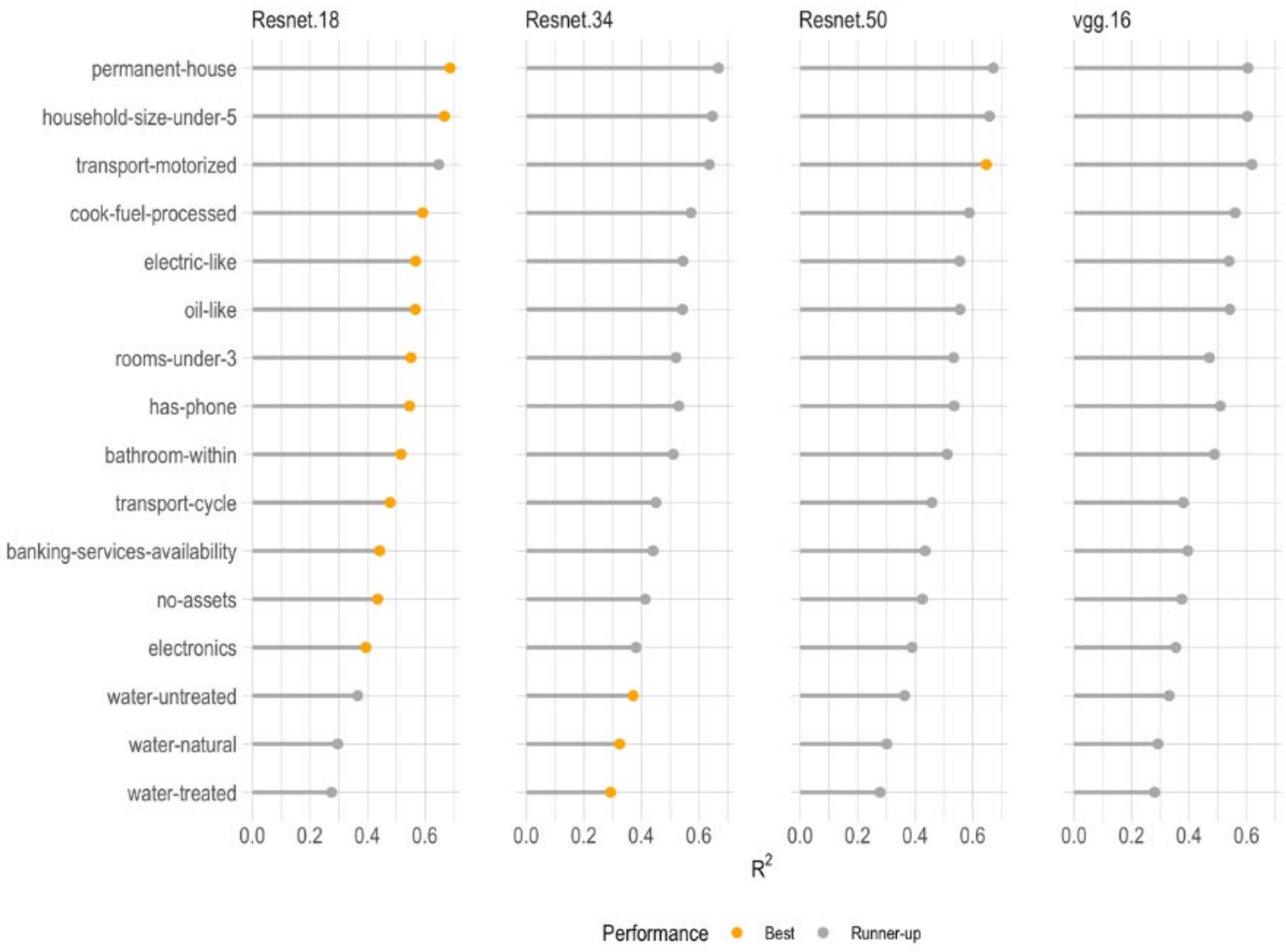

Panel (a): Comparison of different deep-learning models for the prediction of the 2011 village level asset vector using 2011 satellite images. Remotely measuring a 16-dimensional representation of material-living standards using 2011 Census (all algorithms)

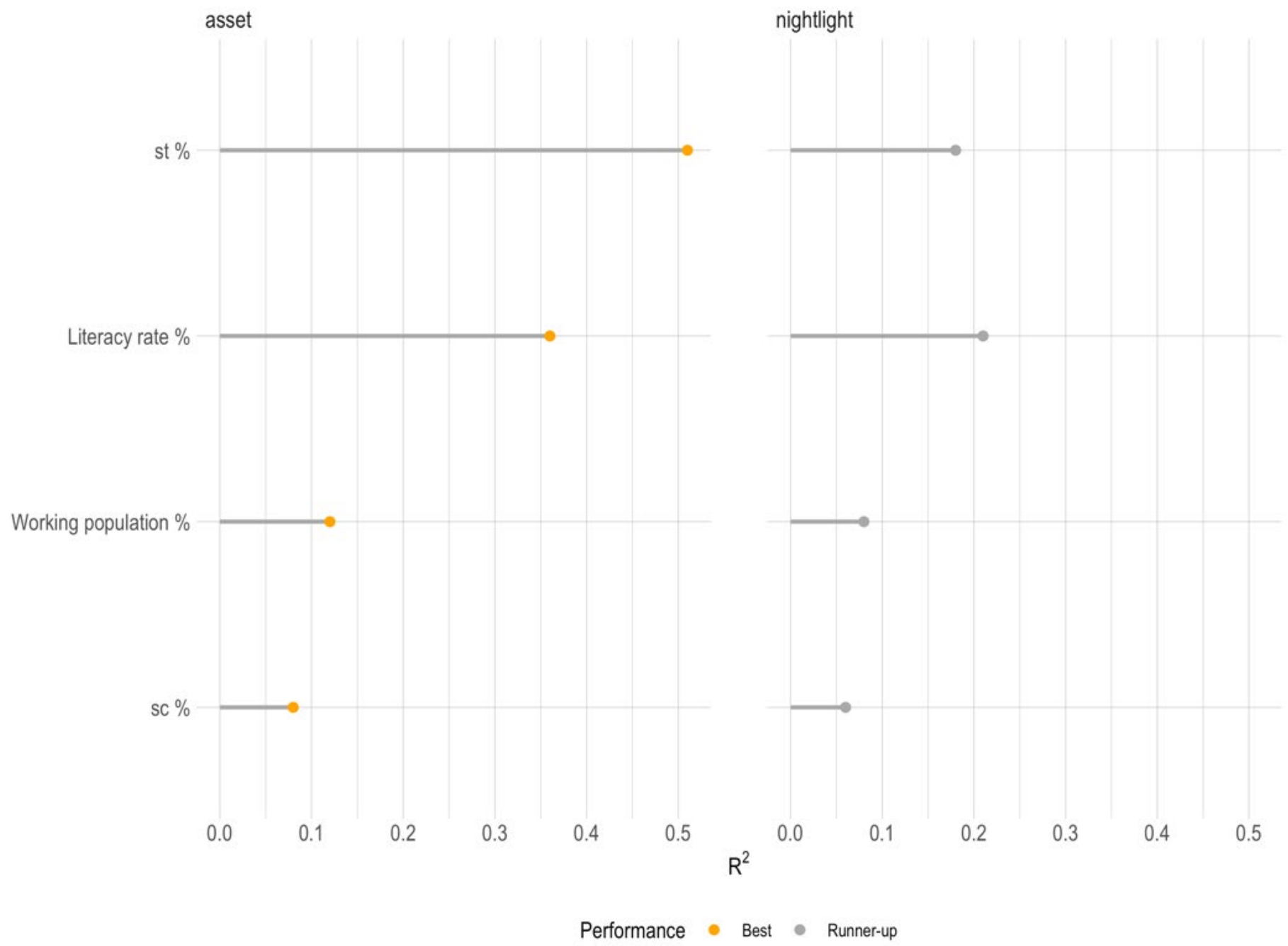

Panel (b): Comparison of transfer learning of summary population and demographic data at the village level using the nightlight and asset models. The Population and Demographic variables are extracted from the Population Census Abstract. The results are based on the 16-dimensional asset model to remotely measuring demographic characteristics with transfer learning.

*Figure 3: Census-based cross-sectional results 2011*

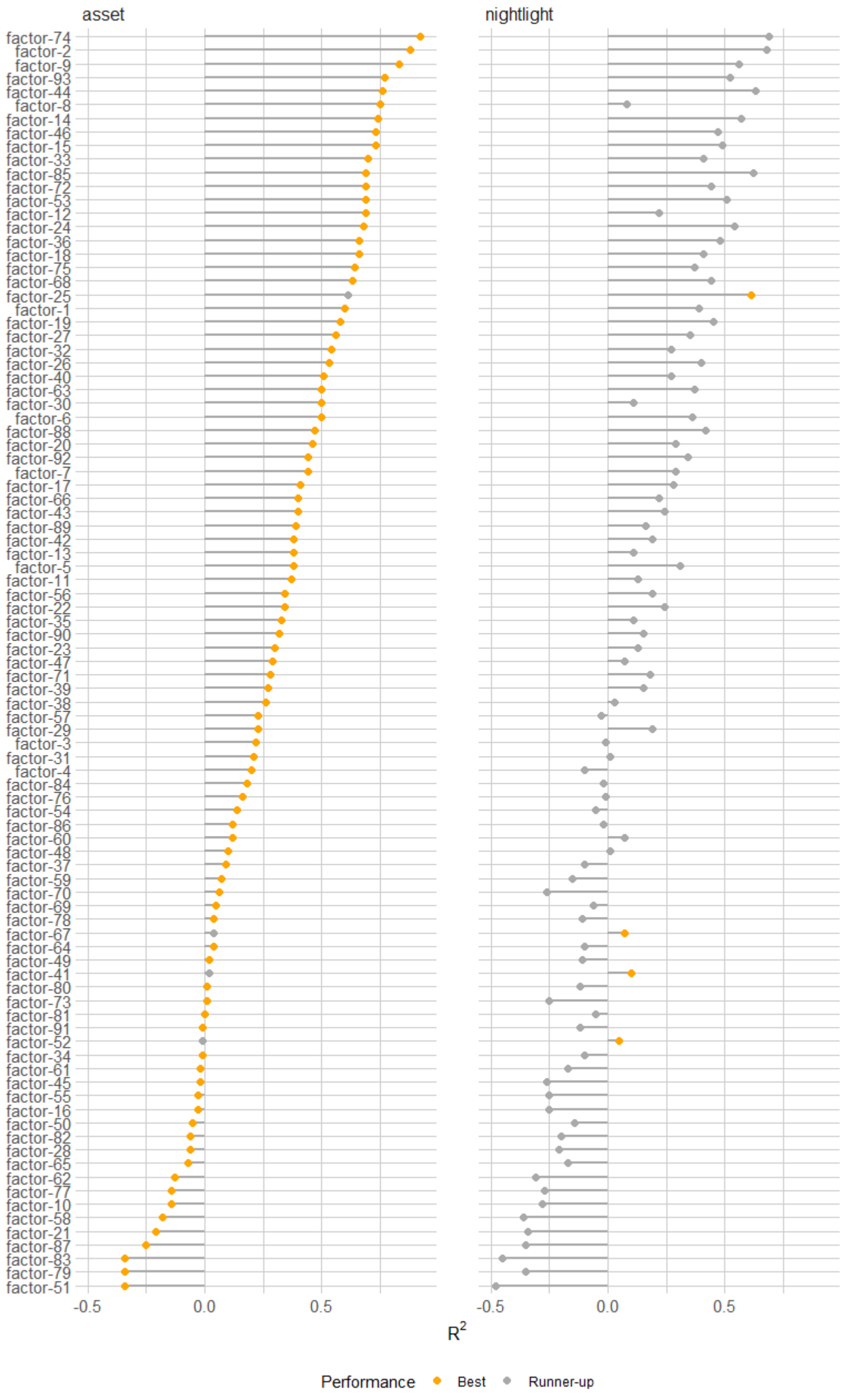

panel a: Comparison of transfer learning of the NFHS-4 variables at the district level using the asset models and 2015 images. The histogram captures the R-squared of the 93 NFHS variables

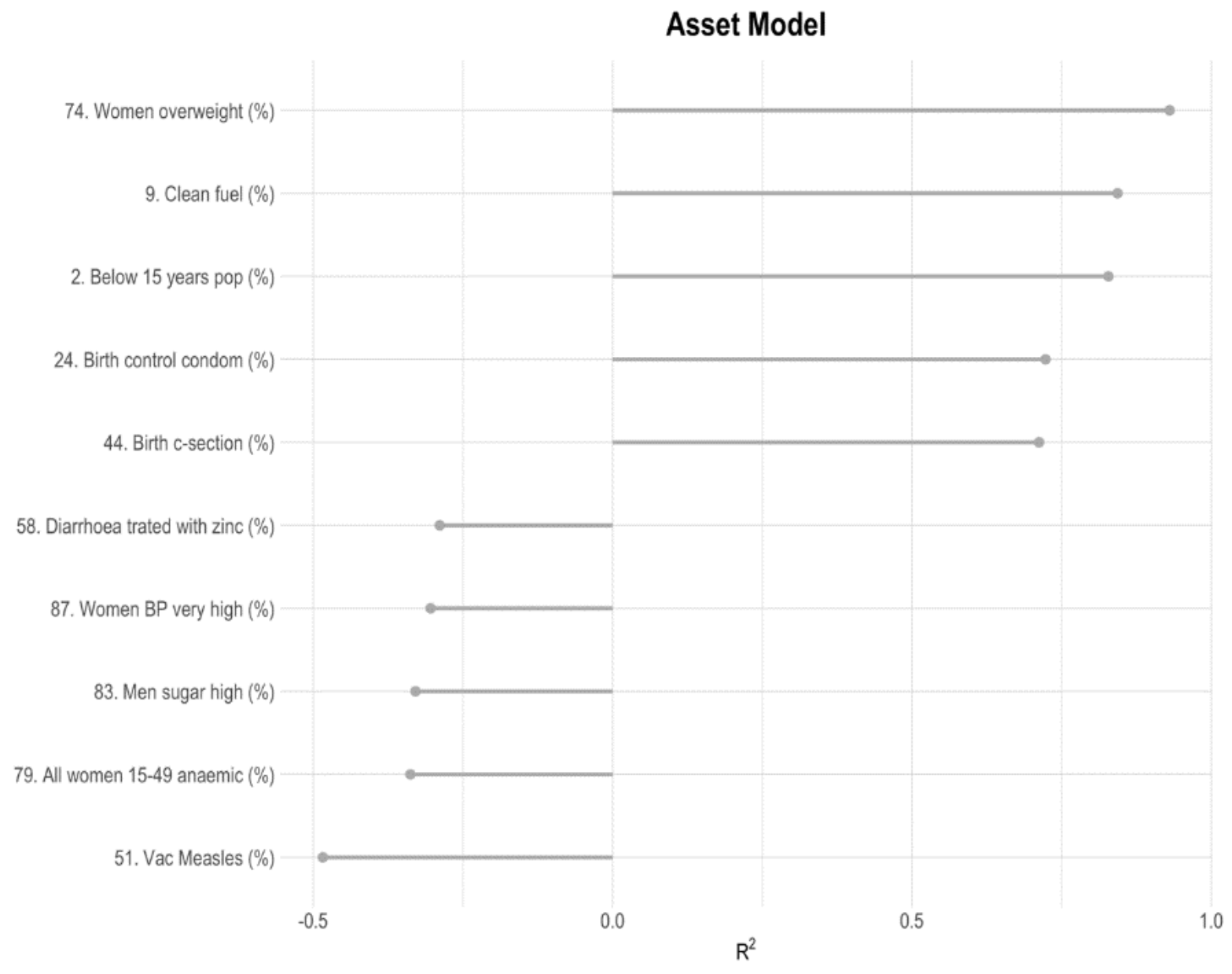

Panel b: the five top-scoring NFHS variables

*Figure 4: NFHS-based cross-sectional results*

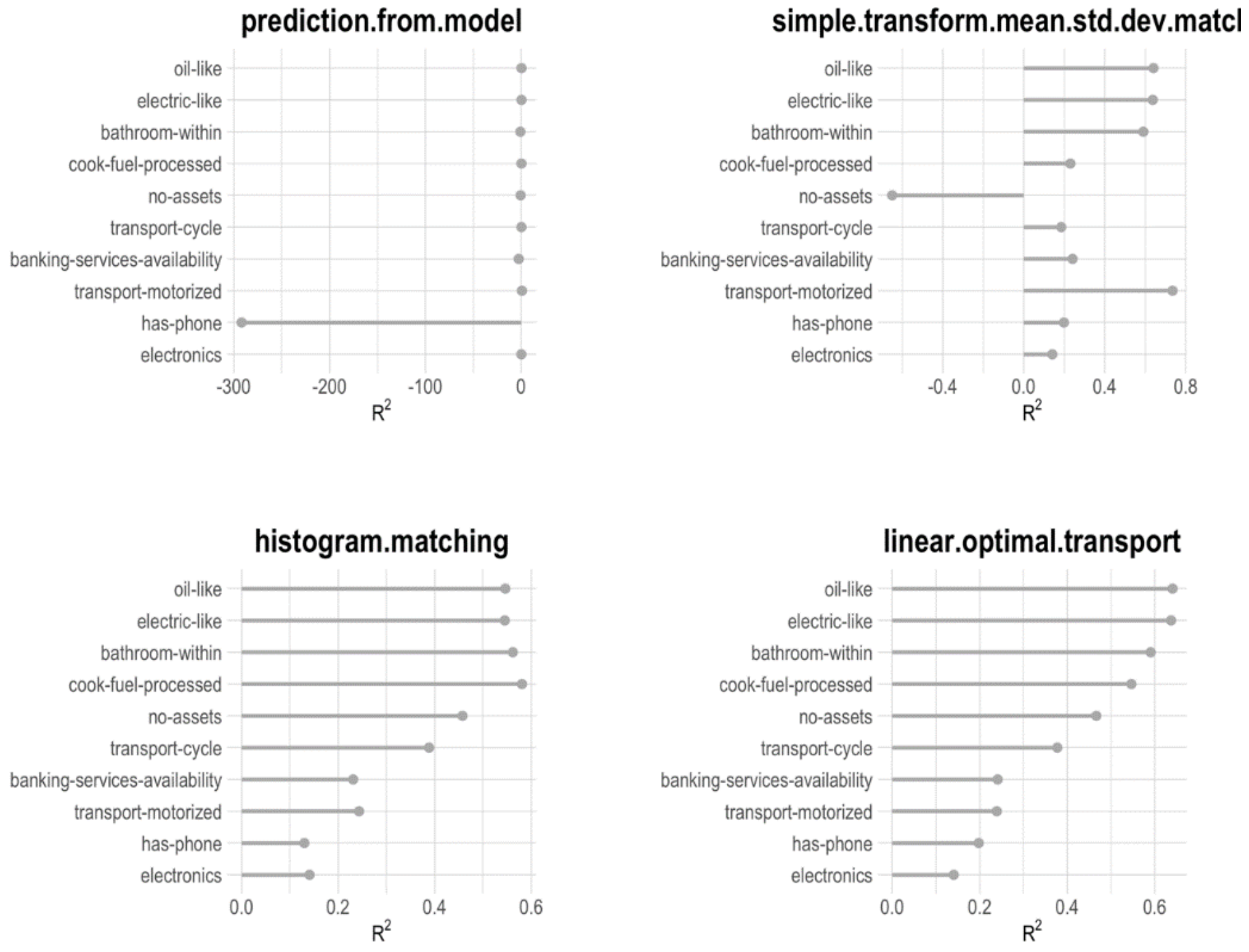

Panel (a) Remotely measuring 2001-census outcomes from 2011-census training and with distribution transformations. The figure shows prediction of the 2001 tehsil level asset vector using 2011 asset model and 2001 images for different temporal transformations of the tehsil level asset distributions

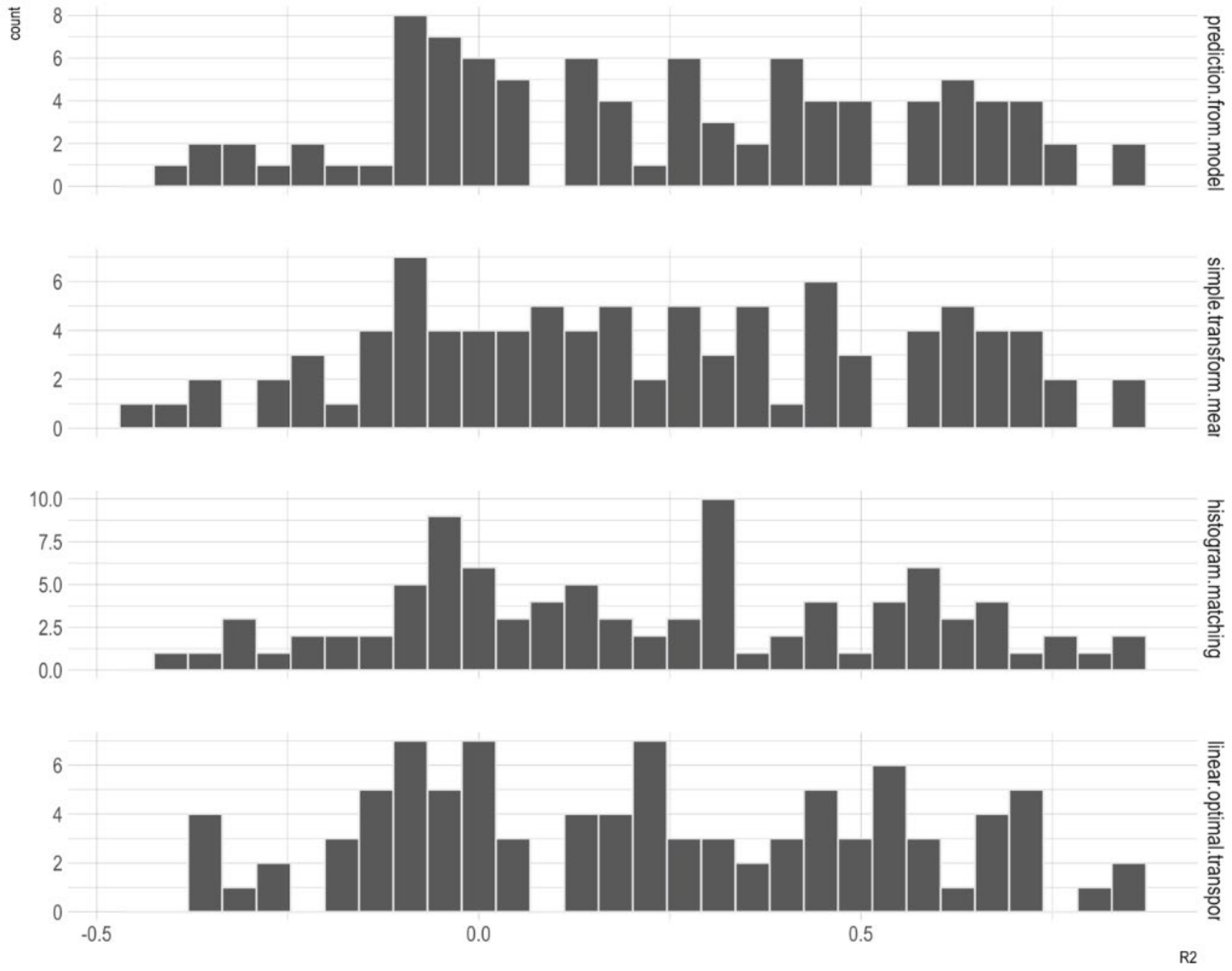

Panel (b) top-five and bottom-five of the 93 NFHS-4. Remotely measuring health outcomes (NFHS-4) from 2011-census training. The figure shows the four histograms for untransformed prediction, simple transform, histogram matching, and linear optimal transport.

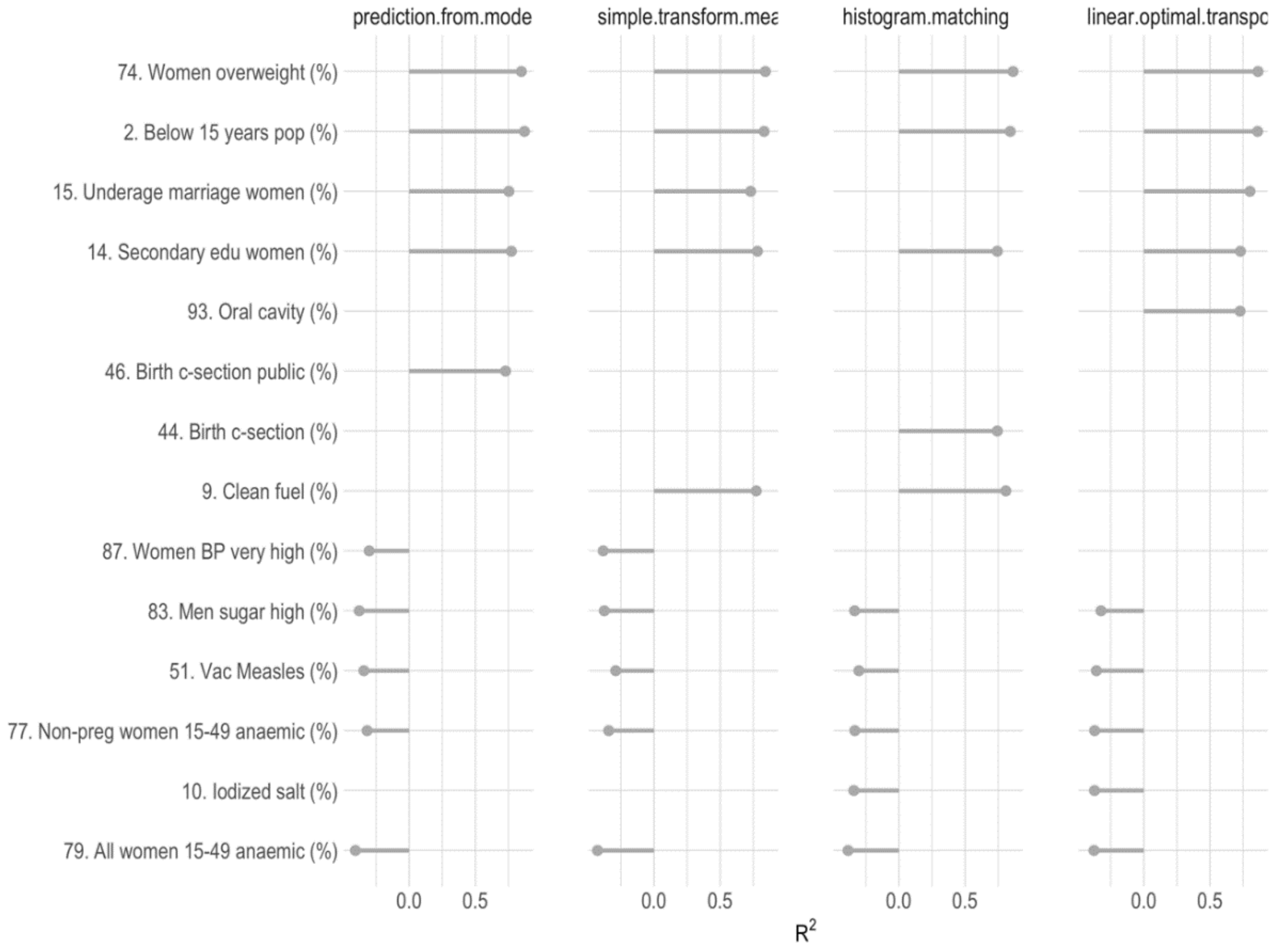

Panel (c) Double-transfer learning (from Census 2011 to NFHS-4, and then to outcomes in NFHS-5) with the selected 3 child outcomes

*Figure 5: Temporal results*

## 6.2 Tables

*Table 1* ***The dimension reduced village level asset vector derived from the Census Data, 2011***

| Variable | Description (percentage of households in the village) | Source (aggregation from columns in the census)[3] |
|---|---|---|
| rooms-under-3 | having less than 3 rooms in the house | [49]+[50]+[51] |
| household-size-under-5 | having less than 5 members in the household | [56]+[57]+[58]+[59] |
| water-treated | having access to water from treated source/covered well/ tube-well | [72]+[74]+[77] |
| water-untreated | having access to water from untreated source/uncovered well | [73]+[75] |
| water-natural | having access to drinking water from ponds/rivers/lakes | [76]+[78]+[79]+[80]+[81] |
| electric-like | having access to electricity from grid/solar | [85]+[87] |
| oil-like | having access to fuel from kerosene/other oil | [86]+[88]+[89] |
| electronics | having possession of radio/transistor/tv/laptop | ([128]+[129]+[130]+[131])/3 |
| has-phone | having possession of land-line/mobile/both | [132]+[133]+[134] |
| transport-cycle | having possession of bicycle | [135] |
| transport-motorized | having possession of motorcycle/scooter/car/jeep | [136]+[137] |
| no-assets | having no asset (cycle/phone etc.) | [139] |

[3] The numbers in the square brackets [] denote the column numbers of the Census data table HLPCA-H14, Percentage of Households to Total Households by Amenities and Assets (India & States/UTs) - Village and Ward Level, Houselisting and Housing Census Data – 2011

| | | |
|---|---|---|
| banking-services-availability | availing banking services | [127] |
| cook-fuel-processed | having possession of LPG/electric stove etc. | [113]+[114]+[115] |
| bathroom-within | having bathroom within premises | [103]+[104] |
| permanent-house | having permanent house | [140] |

*Table 2* ***The dimension reduced village level asset vector derived from the Census Data, 2001***

*Table 3* ***The dimension reduced village level asset vector derived from 2011 and 2001 Census Data***

| **Variable** | **Description (proportion of households in the Tehsil)** | 2011 Census | | 2001 Census | |
|---|---|---|---|---|---|
| | | Aggregation from columns[1] | **Census Data Table** | Aggregation from columns[1] | **Census Data Table** |
| electric-like | having energy from grid/solar | ([9] +[11]) / [8] | HH-7 - Households by Main Source of Lighting | ([3] + [5]) / [2] | H-9 – Distribution of Households by Source of Lighting |
| oil-like | having access to fuel from kerosene/other oil | ([10] +[12]) / [8] | HH-7 - Households by Main Source of Lighting | ([4] + [6]) / [2] | H-9 – Distribution of Households by Source of Lighting |
| electronics | having radio/transistor/tv/computer | (([10] + [11] + [12] + [13])/3 + [20])/ [8] | HH-12 - Number of Households Availing Banking Services and Number of Households Having Each of the Specified Assets | ([4] + [5]) / [2] | H-13 – Number of Households Availing Banking Services and Number of Households Having Each of The Specified Asset |
| has-phone | having telephone/mobile | ([14] + [15] + [16] ) / [8] | HH-12 - Number of Households Availing Banking Services and Number of Households Having Each of the Specified Assets | [6] / [2] | H-13 – Number of Households Availing Banking Services and Number of Households Having Each of The Specified Asset |
| transport-cycle | having bicycle | [17] / [8] | HH-12 - Number of Households Availing Banking Services and Number of Households Having Each of the Specified Assets | [7] / [2] | H-13 – Number of Households Availing Banking Services and Number of Households Having Each of The Specified Asset |
| transport-motorized | having motorcycle/scooter/car/jeep | ([18] + [19]) / [8] | HH-12 - Number of Households Availing Banking Services and Number of Households Having Each of the Specified Assets | ([8] + [9]) / [2] | H-13 – Number of Households Availing Banking Services and Number of Households Having Each of The Specified Asset |
| no-assets | having no assets (cycle/phone etc.) | [21] / [8] | HH-12 - Number of Households Availing Banking Services and Number of Households Having Each of the Specified Assets | [10]/ [2] | H-13 – Number of Households Availing Banking Services and Number of Households Having Each of The Specified Asset |
| banking-services-availability | availing banking services | [9] / [8] | HH-12 - Number of Households Availing Banking Services and Number of Households Having Each of the Specified Assets | [3] / [2] | H-13 – Number of Households Availing Banking Services and Number of Households Having Each of The Specified Asset |
| cook-fuel-processed | having LPG/electric stove etc. | ([14] + [15]) / [9] | HH-10 – Households by Availability of Separate Kitchen and type of Fuel used For Cooking | ([8] + [9]) / [3] | H-10 – Distribution of Households by Availability of Bathroom type of Latrine Within The House and type of Drainage Connectivity for Waste |
| | | | | | Water Outlet Table |
| bathroom-within | having bathroom within premises | [9] / [8] | HH-10 – Households by Availability of Separate Kitchen and type of Fuel used For Cooking | [3] / [2] | H-11 – Distribution of Households by Availability of Separate Kitchen and type of Fuel Used for Cooking |
| | | | | | |
| | | | | | |
| 1[1] The column names in the Census data tables are given in square brackets []. | | | | | |

*Table 4* ***The dimension reduced NFHS district level health vector***

**Table: The dimension reduced NFHS district level health vector**

| Variable | Description |
|---|---|
| factor-1 | Population (female) age 6 years and above who ever attended school (%) |
| factor-2 | Population below age 15 years (%) |
| factor-3 | Sex ratio of the total population (females per 1,000 males) |
| factor-4 | Sex ratio at birth for children born in the last five years (females per 1,000 males) |
| factor-5 | Children under age 5 years whose birth was registered (%) |
| factor-6 | Households with electricity (%) |
| factor-7 | Households with an improved drinking-water source1 (%) |
| factor-8 | Households using improved sanitation facility2 (%) |
| factor-9 | Households using clean fuel for cooking3 (%) |
| factor-10 | Households using iodized salt (%) |
| factor-11 | Households with any usual member covered by a health scheme or health insurance (%) |
| factor-12 | Women who are literate (%) |
| factor-13 | Men who are literate (%) |
| factor-14 | Women with 10 or more years of schooling (%) |
| factor-15 | Women age 20-24 years married before age 18 years (%) |
| factor-16 | Men age 25-29 years married before age 21 years (%) |
| factor-17 | Women age 15-19 years who were already mothers or pregnant at the time of the survey (%) |
| factor-18 | Any method4 (%) |
| factor-19 | Any modern method4 (%) |
| factor-20 | Female sterilization (%) |
| factor-21 | Male sterilization (%) |
| factor-22 | IUD/PPIUD (%) |
| factor-23 | Pill (%) |
| factor-24 | Condom (%) |
| factor-25 | Total unmet need (%) |
| factor-26 | Unmet need for spacing (%) |
| factor-27 | Health worker ever talked to female non-users about family planning (%) |
| factor-28 | Current users ever told about side effects of current method6 (%) |
| factor-29 | Mothers who had antenatal check-up in the first trimester (%) |
| factor-30 | Mothers who had at least 4 antenatal care visits (%) |
| factor-31 | Mothers whose last birth was protected against neonatal tetanus7 (%) |
| factor-32 | Mothers who consumed iron folic acid for 100 days or more when they were pregnant (%) |
| factor-33 | Mothers who had full antenatal care8 (%) |
| factor-34 | Registered pregnancies for which the mother received Mother and Child Protection (MCP) card (%) |
| factor-35 | Mothers who received postnatal care from a doctor/nurse/LHV/ANM/midwife/other health personnel within 2 days of delivery (%) |
| factor-36 | Mothers who received financial assistance under Janani Suraksha Yojana (JSY) for births delivered in an institution (%) |
| factor-37 | Average out of pocket expenditure per delivery in public health facility (Rs.) |
| factor-38 | Children born at home who were taken to a health facility for check-up within 24 hours of birth (%) |
| factor-39 | Children who received a health check after birth from a doctor/nurse/LHV/ANM/ midwife/other health personnel within 2 days of birth (%) |
| factor-40 | Institutional births (%) |
| factor-41 | Institutional births in public facility (%) |
| factor-42 | Home delivery conducted by skilled health personnel (out of total deliveries) (%) |
| factor-43 | Births assisted by a doctor/nurse/LHV/ANM/other health personnel (%) |
| factor-44 | Births delivered by caesarean section (%) |
| factor-45 | Births in a private health facility delivered by caesarean section (%) |
| factor-46 | Births in a public health facility delivered by caesarean section (%) |
| factor-47 | Children age 12-23 months fully immunized (BCG, measles, and 3 doses each of polio and DPT) (%) |
| factor-48 | Children age 12-23 months who have received BCG (%) |
| factor-49 | Children age 12-23 months who have received 3 doses of polio vaccine (%) |
| factor-50 | Children age 12-23 months who have received 3 doses of DPT vaccine (%) |
| factor-51 | Children age 12-23 months who have received measles vaccine (%) |
| factor-52 | Children age 12-23 months who have received 3 doses of Hepatitis B vaccine (%) |
| factor-53 | Children age 9-59 months who received a vitamin A dose in last 6 months (%) |
| factor-54 | Children age 12-23 months who received most of the vaccinations in public health facility (%) |
| factor-55 | Children age 12-23 months who received most of the vaccinations in private health facility (%) |
| factor-56 | Prevalence of diarrhoea (reported) in the last 2 weeks preceding the survey (%) |
| factor-57 | Children with diarrhoea in the last 2 weeks who received oral rehydration salts (ORS) (%) |
| factor-58 | Children with diarrhoea in the last 2 weeks who received zinc (%) |

| | | |
|---|---|---|
| factor-59 | Children with diarrhoea in the last 2 weeks taken to a health facility (%) | |
| factor-60 | Prevalence of symptoms of acute respiratory infection (ARI) in the last 2 weeks preceding the survey (%) | |
| factor-61 | Children with fever or symptoms of ARI in the last 2 weeks preceding the survey taken to a health facility (%) | |
| factor-62 | Children under age 3 years breastfed within one hour of birth9 (%) | |
| factor-63 | Children under age 6 months exclusively breastfed10 (%) | |
| factor-64 | Children age 6-8 months receiving solid or semi-solid food and breastmilk10 (%) | |
| factor-65 | Breastfeeding children age 6-23 months receiving an adequate diet10,11 (%) | |
| factor-66 | Non-breastfeeding children age 6-23 months receiving an adequate diet10,11 (%) | |
| factor-67 | Total children age 6-23 months receiving an adequate diet10,11 (%) | |
| factor-68 | Children under 5 years who are stunted (height-for-age)12 (%) | |
| factor-69 | Children under 5 years who are wasted (weight-for-height)12 (%) | |
| factor-70 | Children under 5 years who are severely wasted (weight-for-height)13 (%) | |
| factor-71 | Children under 5 years who are underweight (weight-for-age)12 (%) | |

| | |
|---|---|
| factor-72 | Women whose Body Mass Index (BMI) is below normal (BMI < 18.5 kg/m2)14 (%) |
| factor-73 | Men whose Body Mass Index (BMI) is below normal (BMI < 18.5 kg/m2) (%) |
| factor-74 | Women who are overweight or obese (BMI √¢‚Ä∞¬• 25.0 kg/m2)14 (%) |
| factor-75 | Men who are overweight or obese (BMI √¢‚Ä∞¬• 25.0 kg/m2) (%) |
| factor-76 | Children age 6-59 months who are anaemic (<11.0 g/dl) (%) |
| factor-77 | Non-pregnant women age 15-49 years who are anaemic (<12.0 g/dl) (%) |
| factor-78 | Pregnant women age 15-49 years who are anaemic (<11.0 g/dl) (%) |
| factor-79 | All women age 15-49 years who are anaemic (%) |
| factor-80 | Men age 15-49 years who are anaemic (<13.0 g/dl) (%) |
| factor-81 | Blood sugar level - high (>140 mg/dl) (%) women |
| factor-82 | Blood sugar level - very high (>160 mg/dl) (%) women |
| factor-83 | Blood sugar level - high (>140 mg/dl) (%) men |
| factor-84 | Blood sugar level - very high (>160 mg/dl) (%) men |
| factor-85 | Slightly above normal (Systolic 140-159 mm of Hg and/or Diastolic 90-99 mm of Hg) (%) women |
| factor-86 | Moderately high (Systolic 160-179 mm of Hg and/or Diastolic 100-109 mm of Hg) (%) women |
| factor-87 | Very high (Systolic >=180 mm of Hg and/or Diastolic >=110 mm of Hg) (%) women |
| factor-88 | Slightly above normal (Systolic 140-159 mm of Hg and/or Diastolic 90-99 mm of Hg) (%) men |
| factor-89 | Moderately high (Systolic 160-179 mm of Hg and/or Diastolic 100-109 mm of Hg) (%) men |
| factor-90 | Very high (Systolic >=180 mm of Hg and/or Diastolic >=110 mm of Hg) (%) men |
| factor-91 | Cervix (%) |
| factor-92 | Breast (%) |
| factor-93 | Oral cavity (%) |